\documentclass[a4paper,10pt]{jpconf}
\usepackage{graphicx}
\usepackage{citesort}
\begin{document}
\title{Continuous-Time Monte Carlo study of the pseudogap Bose-Fermi Kondo model}

\author{J. H. Pixley$^1$, Stefan Kirchner$^{2,3}$, M.~T.~Glossop$^1$ and Qimiao Si$^1$}

\address{$^1$ Department of Physics \& Astronomy, Rice University,
Houston, Texas, 77005, USA}
\address{$^2$ Max Planck Institute for the Physics of Complex Systems,
01187 Dresden, Germany}
\address{$^3$ Max Planck Institute for Chemical Physics of Solids,
01187 Dresden, Germany}

\ead{jp11@rice.edu}

\begin{abstract}
We study the pseudogap Bose-Fermi Anderson model with a continuous-time quantum Monte Carlo (CT-QMC) method.  We discuss some delicate aspects of the transformation from this model to the Bose-Fermi Kondo model.  We show that the CT-QMC method can be used at sufficiently low temperatures to access the quantum critical properties of these models. 
\end{abstract}


Over the past decade intermetallic compounds have served as model systems  to study instabilities  of metallic magnets near zero temperature~\cite{Sachdev,Loehneysen.07,Gegenwart.08}. In particular the quantum critical properties found in
several heavy fermion compounds seem to be beyond the Ginzburg-Landau paradigm
of criticality~\cite{Aronson.95,Schroeder.00,Paschen.04,Friedemann.10}. 
A natural
explanation for the
new type of criticality
invokes inherently-quantum critical modes in addition to the
gapless order
parameter fluctuations. In the case of
heavy fermion compounds,
the additional critical mode has been identified with the destruction 
of the Kondo effect~\cite{Coleman.01,Si.01,Si.03}.
Since the destruction of the Kondo effect is local in space 
it can be systematically studied
in simplified, local models. 
In these quantum impurity systems the 
critical state only appears at the (spatial) boundary of suitable
hosts.

In the present article, we study the Kondo-destroying quantum critical point
in the pseudogap Bose-Fermi Kondo model (PBFKM) with Ising anisotropy,
 defined as
\begin{eqnarray}
\label{Eq:PBFKM}
H_{\mbox{\tiny PBFKM}} &=&\, \sum_{k, \sigma} \epsilon_k^{} c_{k,\sigma}^{\dag}c_{k,\sigma}^{} + \sum_{q} \omega_q \phi_{q}^{\dag}\phi_{q}
+ J{\bf S}\cdot \sum_{k,k',\sigma,\sigma'}  c_{k,\sigma}^{\dag}\frac{\vec{\sigma}}{2}c_{k',\sigma'}+g S_z \sum_q (\phi_{q}^{\dag}+\phi_{-q}^{}).
\label{pbfkm}
\end{eqnarray}
Where $\epsilon_k$, $\omega_q$ are the fermonic and bosonic bath dispersions, $J$ is the Kondo coupling between the spin of the conduction electrons and the spin of the impurity, $\vec{\sigma}$ is a vector of Pauli spin matrices and $g$ is the coupling between the $z$-component of the impurity spin and the bosonic bath.  We take a pseudogap density of states (DOS) for the conduction electrons, $\rho_c(\epsilon) \propto |\epsilon|^r$ for $|\epsilon| < D$ and $0<r<1/2$ where we have taken the Fermi energy to be zero, and a sub-ohmic density of states for the bosons, 
$\rho_B(\omega) =\sum_q  [\delta(\omega-\omega_q)-\delta(\omega+\omega_q)]\propto \mathrm{sgn}(\omega)|\omega|^{\alpha}$ 
up to a cutoff $\Lambda$.
A perturbative renormalization group study has been used carried out for the Bose-Fermi Kondo model
with non-zero conduction-electron DOS ($r=0$) in both the Ising and continuous
spin symmetry cases \cite{zhu.02}. For the pseudogapped case ($r \ne 0$), 
such a study has been possible only for the continuous spin symmetry case \cite{Vojta.04}.
The numerical renormalization group method, on the other hand, has been used to study the model 
in the Ising limit \cite{Glossop.08}. 
Eq.~(\ref{pbfkm}) has 
as particular limits the pseudogap Kondo and the Bose-Fermi Kondo models,
which have recently been
studied
using related methods ~\cite{Kirchner.08,Glossop.10}.  
Despite the seeming simplicity of  Eq.~(\ref{Eq:PBFKM}), it is hard
to study  the quantum critical properties of the PBFKM directly. 
In particular, the quantum-relaxational regime ($\hbar \omega < k_BT$) is difficult to address theoretically. The quantum-relaxational regime of the pseudogap Kondo model  has been
studied previously~\cite{Glossop.10}, by applying a 
CT-QMC algorithm~\cite{Werner.06,Prokofev.98,Rubtsov.05} to the pseudogap Anderson model whose low-energy sector in the local moment regime can be mapped onto  the pseudogap Kondo model via a Schrieffer-Wolff (SW) transformation.
Since the Anderson model only involves fermions, standard perturbative expansion methods can be applied to it.
In the CT-QMC approach, the full perturbative expansion in terms of the 
hybridization between conduction and localized fermions is then 
sampled  stochastically using a Monte Carlo algorithm~\cite{Werner.06,Prokofev.98,Rubtsov.05}.
The corresponding Bose-Fermi Anderson model is
\begin{eqnarray}
H &=& \sum_{k, \sigma} \epsilon_k c_{k,\sigma}^{\dag}c_{k,\sigma} + \epsilon_d(n_{\uparrow} + n_{\downarrow}) + U n_{\uparrow}n_{\downarrow} 
 + \sum_{k, \sigma}( V_k d_{\sigma}^{\dag}c_{k,\sigma} + V_k^*c_{k,\sigma}^{\dag}d_{\sigma}) 
 \nonumber \\
&&+ \sum_{q} \omega_q \phi_{q}^{\dag}\phi_{q} + g\frac{(n_{\uparrow} - n_{\downarrow})}{2}\sum_{q}(\phi_{q}^{\dag} + \phi_{-q}),
\label{Hamiltonian_1}
\end{eqnarray}
where 
$\epsilon_d$ is the energy level of the impurity, 
$U$ is the
on-site
interaction ,
$n_{\sigma}=d^{\dag}_{\sigma}d_{\sigma}$, 
$V_k$ is the hybridization of the impurity 
with the conduction electrons, 
and $S_z = \frac{1}{2} (n_{\uparrow} - n_{\downarrow})$.
The CT-QMC method has been extended to treat ohmic bosonic baths
coupled to the charge of the impurity by invoking a Firsov-Lang (FL)
transformation~\cite{Werner.07,werner_millis.10}.  Here we extend
this approach to treat a sub-ohmic bosonic bath that couples 
to the spin of the impurity and
explore the possibility of using it to access the quantum critical 
properties.
As it turns out, the generators of the
FL and SW transformations do not commute,
raising the important question as to
which is the proper order of applying the two transformations.

We focus on the particle-hole symmetric case, $U = -2\epsilon_d$.  First, we perform the FL transformation to eliminate the term linear in $\phi$
exactly.  We choose a generator $S_{FL}= g S_z \sum_q \frac{1}{\omega_q} ( \phi^{\dag}_q - \phi_{-q})$ and the transformed Hamiltonian $\tilde{H} = e^{S_{FL}}He^{-S_{FL}}$
is
\begin{equation}
\tilde{H} = \sum_{k, \sigma} \epsilon_k c_{k,\sigma}^{\dag}c_{k,\sigma} + \tilde{\epsilon_d}(n_{\uparrow} + n_{\downarrow}) + \tilde{U} n_{\uparrow}n_{\downarrow}  
 + \sum_{k, \sigma}( V_k \tilde{d}_{\sigma}^{\dag}c_{k,\sigma} + V_k^*c_{k,\sigma}^{\dag}\tilde{d}_{\sigma}) + \sum_q \omega_q \phi^{\dag}_q\phi_q, 
\label{H_2}
\end{equation}
where $\tilde{d}_{\sigma}^{\dag}= d_{\sigma}^{\dag}\mathrm{exp}(\frac{\sigma g}{2}\sum_q\frac{1}{\omega_q}(\Phi_{q}^{\dag} - \Phi_{-q}))$, $\sigma = \pm 1$ for $\uparrow$/$\downarrow$, $\tilde{U} = U + \frac{1}{2}g^2\sum_q\frac{1}{\omega_q}$, and $\tilde{\epsilon}_d = -\frac{\tilde{U}}{2}$. 
Note that $\tilde{n}_{\sigma}=n_{\sigma}$ and the transformation does not destroy particle-hole symmetry.  


A modified 
SW transformation 
~\cite{S_W} 
is used
to eliminate the hybridization term in Eq.~(\ref{H_2}).  
We only consider the 
Kondo limit, and for simplicity neglect the $k$ dependence 
of $V_k = V$.
The generator $S_{SW}$ is
standard
but with $U, \epsilon_d, d, d^{\dag}$ replaced by $\tilde{U}, \tilde{\epsilon_d}, \tilde{d}, \tilde{d}^{\dag}$, namely $S_{SW}=\sum_{k,\sigma}V\Big( \frac{1-n_{-\sigma}}{\tilde{\epsilon_d}-\epsilon_k}+\frac{n_{-\sigma}}{\tilde{\epsilon_d}+\tilde{U}
-\epsilon_k}\Big)(\tilde{d}_{\sigma}^{\dag}c_{k,\sigma} - c_{k,\sigma}^{\dag}\tilde{d}_{\sigma})$.  Writing the Hamiltonian in Eq.~(\ref{H_2}) as $\tilde{H} = H_0 + H_b + \tilde{H}_h$ where
$H_0=\tilde{\epsilon_d}(n_{\uparrow} + n_{\downarrow}) + \tilde{U} n_{\uparrow}n_{\downarrow} + \sum_{k, \sigma} \epsilon_k c_{k,\sigma}^{\dag}c_{k,\sigma}$, $H_b = \sum_q \omega_q \phi_q^{\dag}\phi_q$ and $\tilde{H}_h=\sum_{k,\sigma}V(\tilde{d}_{\sigma}^{\dag}c_{k,\sigma} + c_{k,\sigma}^{\dag}\tilde{d}_{\sigma})$, we have $H^{\prime} = e^{S_{SW}}\tilde{H}e^{-S_{SW}} \approx H_0 + H_b + [S_{SW},H_b] + \frac{1}{2}[S_{SW},\tilde{H}_h]$.
Projecting out unoccupied and doubly occupied states we arrive at
\begin{eqnarray}
H^{\prime} &=& \sum_{k, \sigma} \epsilon_k c_{k,\sigma}^{\dag}c_{k,\sigma} + \sum_q \omega_q \phi^{\dag}_q \phi_q + \sum_{k, k', \sigma} (\frac{1}{2}\tilde{W}_{k,k'} + \frac{1}{4}\tilde{J}_{k,k'})c_{k,\sigma}^{\dag}c_{k',\sigma}
\nonumber \\
&-& \sum_{k, k^{\prime}} \tilde{J}_{k, k^{\prime}}( \frac{1}{2}(s^+_{k,k^{\prime}}\tilde{S}^{-} + s^-_{k,k^{\prime}}\tilde{S}^{+})+s^z_{k,k^{\prime}}S^z),
\label{H_SW}
\end{eqnarray}
where $\tilde{W}_{k,k^{\prime}}=V^2(\frac{1}{\epsilon_k - \tilde{\epsilon}_d} + \frac{1}{\epsilon_{k'} - \tilde{\epsilon}_d})$,
$\tilde{J}_{k,k^{\prime}}=V^2(\frac{1}{\epsilon_k -\tilde{\epsilon}_d - \tilde{U} } + \frac{1}{\epsilon_{k^{\prime}}-\tilde{\epsilon}_d - \tilde{U} }-\frac{1}{\epsilon_k-\tilde{\epsilon}_d  } - \frac{1}{\epsilon_{k^{\prime}}-\tilde{\epsilon}_d} )$ is the Kondo coupling, $\tilde{S}^+ = S^+\mathrm{exp}(g\sum_q\frac{1}{\omega_q}(\Phi_{q}^{\dag} - \Phi_{-q}))$, $\tilde{S}^- = S^-\mathrm{exp}(-g\sum_q\frac{1}{\omega_q}(\Phi_{q}^{\dag} - \Phi_{-q}))$, $\vec{S} = \frac{1}{2}\sum_{\alpha,\beta} d^{\dag}_{\alpha}\vec{\sigma}_{\alpha, \beta}d_{\beta}$ and $\vec{s}_{k,k'} = \frac{1}{2}\sum_{\alpha,\beta} c^{\dag}_{k,\alpha}\vec{\sigma}_{\alpha, \beta}c_{k'\beta}$.
The third term in equation~(\ref{H_SW}) represents a potential scattering of the conduction electrons, 
and the 
last is the Kondo term, but with renormalized impurity spin flip operators due to the presence of the bosonic bath. 
$\tilde{J}_{k,k'}$ differs from the standard
expression in that $U$ and $\epsilon_d$ is replaced by $\tilde{U}$ and $\tilde{\epsilon}_d$.  


We will now discuss the opposite order of transformations, namely, $H'' = e^{S_{FL}}e^{S_{SW}}He^{-S_{SW}}e^{-S_{FL}}$.  Applying the SW transformation, projecting out charge fluctuations and then applying the FL transformation arrives at equation (\ref{H_SW}), however with $\tilde{W}_{k,k^{\prime}}$, $\tilde{J}_{k,k^{\prime}}$ replaced by $W_{k,k^{\prime}}$, $J_{k,k^{\prime}}$. We see that applying first the SW transformation, which is not 
exact, completely ignores the bosonic baths' influence on the charge degrees of freedom of the impurity.  Whereas applying the FL transformation first, which is
exact, correctly captures 
the bosonic baths' influence on the Anderson model
which lowers the Kondo coupling.  
The non-commutativity of the two transformations quantitatively
affects the effective Kondo scale at the quantum critical point, but it does not change the universal scaling behavior of the quantum critical properties because the critical value of the Kondo coupling is not universal.


\begin{figure}[t!]
\begin{minipage}[b]{10pc}
\includegraphics[width=1.5in, height=2.0in, angle= -90]{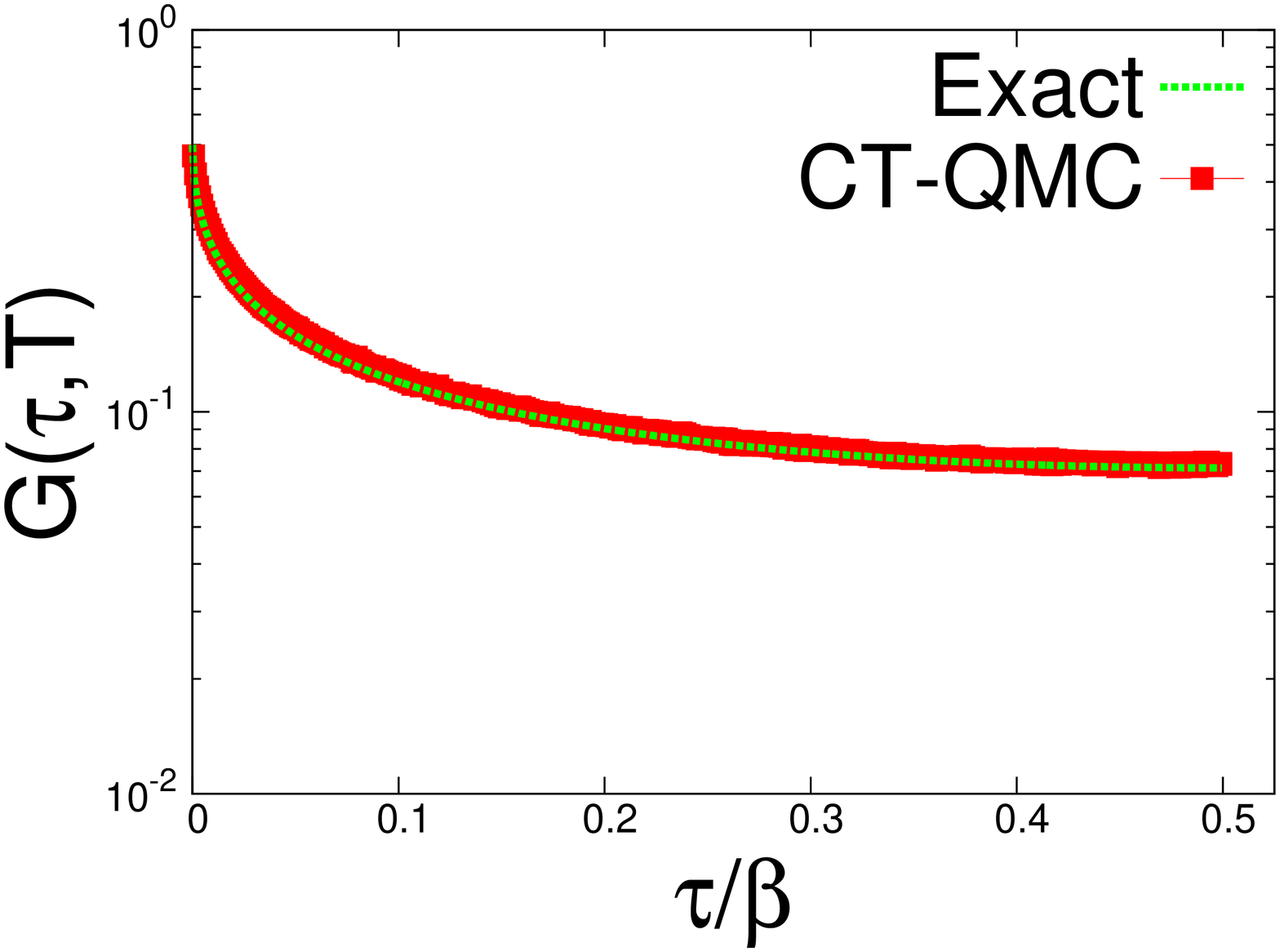}
\end{minipage}\hspace{2.0pc}%
\begin{minipage}[b]{10pc}
\includegraphics[width=1.5in, height=2.0in, angle= -90]{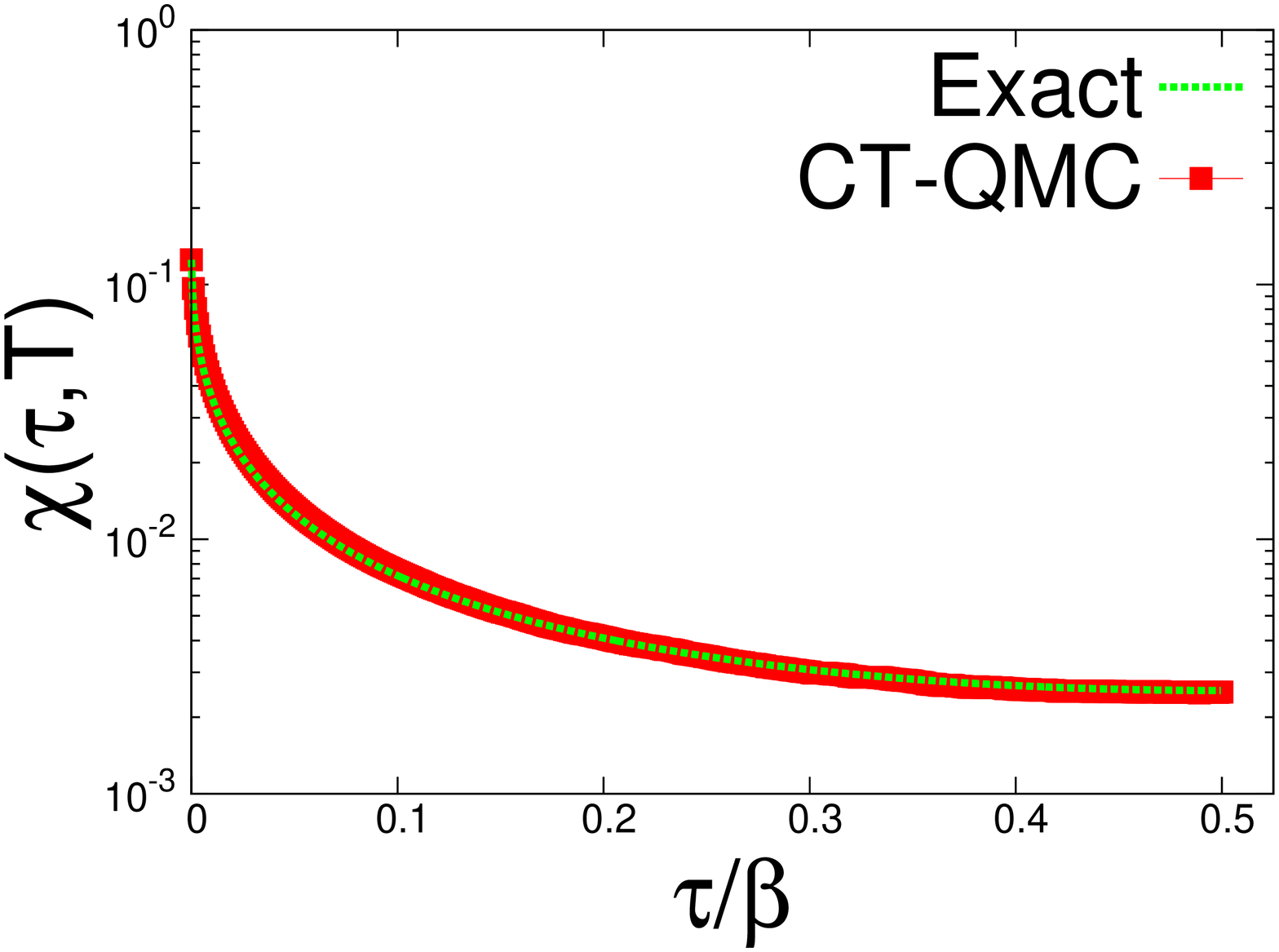}
\end{minipage} \hspace{2pc}
\begin{minipage}[t]{14pc}\caption{\label{fig:Figure2}The single particle Green function $G(\tau,T)$ and the spin susceptibility $\chi(\tau,T)$ for $U=g=0$, $\Gamma = 0.1D$ and $r=0.4$ at $T=6.6\times 10^{-4}D$.  Comparing the exact calculation and the CT-QMC result we see agreement for both $G$ and $\chi$.}
\end{minipage}

\end{figure}

As a check on the CT-QMC approach we first compare the
single particle Green function and the spin susceptibility 
for the numerical result with $U=g=0$ to the analytic result.
For $g=0$ the bosonic bath decouples from the problem and can be ignored, and taking $U=0$ reduces the Hamiltonian in Eq.~(\ref{Hamiltonian_1}) to the resonant level model with a pseudogap.  The impurity single particle Green function, $\langle \hat{T}_{\tau} d(\tau)d^{\dag}(0) \rangle $ is then $G(\omega) =  (\omega - \epsilon_d - \sum_k \frac{|V_k|^2}{\omega - \epsilon_k})^{-1}$. 
Using $V_k = V$ and taking the infinite bandwidth limit we can perform the sum over $k$
~\cite{Buxton.98}.  We obtain $G(\omega) =  (\omega - \epsilon_d - \Sigma(\omega))^{-1}$, where $\mathrm{Re}\Sigma(\omega) = -\Gamma(\omega)\mathrm{tan}(\frac{\pi r}{2})\mathrm{sgn}(\omega)$, $\mathrm{Im}\Sigma(\omega) = -\Gamma(\omega)$ and we have defined the dynamic hybridization function to be $\Gamma(\omega) = \pi |V|^2\rho_0|\omega|^r$.  The imaginary time Green function can then be obtain by Fourier transform, $G(\tau, \beta)  =  \int_{-\infty}^{\infty} \,\frac{\mathrm{d}\omega}{\pi} \frac{e^{-\tau \omega}}{e^{-\beta \omega}+1}\mathrm{Im}(G(\omega + i0^{+}))$ and the local spin susceptibility can be constructed $\chi(\tau,\beta) = -\frac{1}{2}G(\tau,\beta)G(-\tau,\beta)$.  
As seen in figure~\ref{fig:Figure2}
we obtain quantitative agreement, within numerical accuracy, in the long time behavior for both $G(\tau,\beta)$ and $\chi(\tau,\beta)$.

\begin{figure}[t!]
\includegraphics[width=10pc, angle=-90]{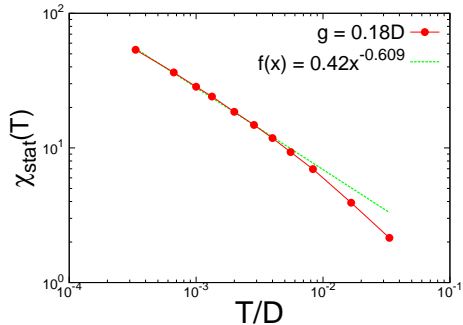}\hspace{2pc}%
\begin{minipage}[t]{16pc}\caption{\label{fig:Figure1}Static susceptibility in the vicinity of the quantum critical point, $g_c \approx 0.18D$.  For $r=0$ the Kondo temperature is, $T_K \approx 0.029D$.  Well below $T_K$ we see divergence of the static susceptibility as $\chi_{\mathrm{stat}} \sim T^{-0.609}$.}
\end{minipage}
\end{figure}

We now turn to the quantum-critical properties of the PBFKM defined in Eq.~(\ref{Hamiltonian_1}) by measuring the static spin susceptibility, described below. 
In what follows we fix $r=0.4$ and $\alpha=0.6$.  After the (FL) transformation, the Hamiltonian in Eq.~(\ref{H_2}) can be expanded in the hybridization term. We use the CT-QMC algorithm
to calculate the partition function, the single particle Green function and the local spin susceptibility.
Fixing $U=0.025D$ and varying $g$ we can tune the model 
 to a quantum critical point (QCP).  Within the CT-QMC approach we measure the local spin susceptibility $\chi(\tau, \beta) = \langle T_{\tau} S_z(\tau) S_z(0) \rangle$ and then calculate the static susceptibility $\chi_{\mathrm{stat}}(\beta) = \int_0^{\beta} d \tau \, \chi(\tau,\beta)$, where we have set the Lande g-factor and Bohr magneton to unity.  For small $g$, the finite $U$ gives rise to a Kondo screened local moment; the static susceptibility approaches a constant for temperatures well below the Kondo temperature, $T_K$.  For large $g$, the impurity spin decouples from the conduction band and follows the fluctuations of the bosonic bath; the static susceptibility takes the Curie-Weiss form, $\chi_{\mathrm{stat}}(T) \sim T^{-1}$.  At the QCP, the bosonic bath acts to decohere and destroy the Kondo effect \cite{Kirchner.08}.  Consequently, at the QCP the scaling of $\chi_{\mathrm{stat}}(T)$ acquires an anomalous  exponent 
$\chi_{\mathrm{stat}}(T)\sim T^{-\alpha}$ 
for temperatures well below $T_K$.  As seen in figure~\ref{fig:Figure1}, using the CT-QMC approach we obtain $g_c \approx 0.18D$
and in the vicinity of the quantum critical point, $\chi_{\mathrm{stat}}(T)\sim T^{-x}$ with $x = 0.609$.
Our calculated exponent agrees with 
the numerical renormalization group result within numerical accuracy~\cite{Glossop.08}; the same exponent is also expected
in related pseudogap Bose-Fermi Kondo model with continuous spin symmetry  \cite{Vojta.04} or the Bose-Fermi Kondo model
with Ising symmetry but with $r=0$ \cite{zhu.02}.


In conclusion, we have shown that  the low energy properties of the pseudogap Bose-Fermi Kondo model can be addressed within
a 
continuous-time quantum Monte Carlo approach. 
We have demonstrated that this approach correctly reproduces the exactly solvable limit 
of the pseudogap resonant level model,
and been able to determine the critical behavior of the static local spin susceptibility in
an interacting case.

This work has been supported by NSF (Grant No. DMR-1006985),
the Robert A. Welch Foundation (Grant No. C-1411),
and the W. M. Keck Foundation.  The calculations were performed on the Rice Computational Research Cluster
funded by the NSF and a partnership between Rice University, AMD and Cray.



\end{document}